# Giant Shifts of Crystal-field Excitations in $ErFeO_3$ Driven by Internal Magnetic Fields

*Joel O'Brien,[1] Guochu Deng,[2*] Xiaoxuan Ma,[3] Zhenjie Feng,[3] Wei Ren,[3] Shixun Cao,[3†] Dehong Yu,[2] Garry J McIntyre,[2] Clemens Ulrich[1]*

[1] *School of Physics, The University of New South Wales, Kensington, NSW 2052, Australia*

[2] *Australian Centre for Neutron Scattering, Australian Nuclear Science and Technology Organisation, New Illawarra Road, Lucas Heights, NSW 2234, Australia*

[3] *Department of Physics, International Centre of Quantum and Molecular Structures and Materials Genome Institute, Shanghai University, 99 Shangda Road, Shanghai 200444, People's Republic of China*

**Abstract**

Due to the complex interactions between rare-earth elements and transition metals, as well as/or themselves, rare-earth transition-metal oxides are likely to exhibit highly intriguing and novel magnetic structures and dynamic behaviours. Rare-earth elements in these compounds frequently demonstrate unusual behaviours in their crystal-field (CF) excitations, which necessitate thorough research for in-depth comprehensions. When cooling from 10 K to 1.5 K via the magnetic ordering temperature of $Er^{3+}$ at 4.1 K, we observed a significant energy shift of the low-lying CF excitation of $Er^{3+}$ in $ErFeO_3$ from 0.35 meV to 0.75 meV utilizing the inelastic neutron-scattering technique. A sound CF model was proposed for $Er^{3+}$ in $ErFeO_3$ by fitting to the observed CF excitation peaks, which enables to explain all the observed experimental results in a very consistent manner. According to the model, the ground crystal field level of $Er^{3+}$, which corresponds to the lowest Kramers doublet supposed to be at zero energy, has been shifted by the internal magnetic fields induced by both $Er^{3+}$ and $Fe^{3+}$ spin orders below and above the $Er^{3+}$ ordering temperature, respectively. Additional measurements in various magnetic fields offer compelling evidence in favour of this hypothesis. The measured external field dependence of the CF excitation energy led to the derivation of the internal field of $Er^{3+}$ as 0.33 meV, which is strongly corroborated by theoretical modelling. Additionally, the effective *g*-factor for $Er^{3+}$ in $ErFeO_3$ showed an exceptionally significant anisotropy.

## I. INTRODUCTION

Transition-metal oxides (TMO) exhibit a wide range of intriguing properties, including high-temperature superconductivity, enormous magnetoresistance, frustrated magnetism, multiferroicity, etc., as a result of the interaction and competition between the spin, charge, and orbital degrees of freedom. The complex magnetic interactions in these compounds leads to exotic magnetic ground states such as incommensurate magnetic structures[1] and frustrated quantum spin liquids.[2] A rare-earth TMO perovskite consists of a network of corner-sharing $MO_6$ octahedra with a rare-earth ion filling the cuboctahedra in-between. The magnetic properties of rare-earth TMO perovskites are primarily dominated by the exchange interactions within the transition-metal sublattice in the intermediate and high-temperature range. While rare-earth ions start to play significant roles in the low-temperature range, rare-earth spins start to interact with the transition-metal spins and induce extra magnetic polarization in the magnetic order on the transition-metal sublattice, which may cause magnetic phase transitions in some cases. Certain rare-earth ions tend to form long-range magnetic orders at very low temperatures as well.[3,4]

In rare-earth compounds, the symmetric Coulomb potential generated by the surrounding cations causes lifting of the degeneracy of the rare-earth 4*f* electron levels. The crystal-field (CF) effect of rare-earth ions can be considered a perturbation to the lowest-energy multiplets of 4*f* ions.[5] The corresponding CF excitations in these compounds fall into an energy range from a few meVs to hundreds of meVs, comparable with the cold- or thermal-neutron energy, and thus can be measured using inelastic neutron scattering[6-8], as an alternative method to solid-state spectroscopies such as electron spin resonance[9] and infrared spectroscopy[10-12]. CF excitations are generally considered local phenomena (nondispersive) due to the weak and almost negligible interaction between the neighbouring rare-earth ions. Due to the Zeeman effect, CF excitation energies could vary by applying external magnetic fields or inducing internal fields by forming long-range magnetic ordering. Boothroyd *et al.* suggested that CF energy shifts could even be used as a sensitive probe to determine the local electrical or magnetic fields on the rare-earth sites.[13]

In the past few decades, numerous experiments have been performed to study CF excitations in compounds with different local symmetries. In the CF excitations of $Ce^{3+}$ in the $LaAl_2$ matrix and other lanthanide-alloy compounds, Loewenhaupt *et al.* [6,8] discovered temperature-dependent energy shifting and linewidth narrowing. Metallic rare-earth systems demonstrated the linewidth change driven by temperature, which Becker et al. attributed to the damping effect induced by conduction electron-hole excitations.[14] Below their superconducting phase transitions of $La_{1-x}Tb_xAl_2$[7] and $PrOs_4Sb_{12}$,[15] the abrupt reduction of CF linewidth was seen and attributed to the suppressed electron scattering upon the opening of the superconducting gap. A similar linewidth shrinkage in $Pr_2O_2M_2OSe_2$ (M = Mn, Fe) below their magnetic ordering temperatures was attributed to the zone-centre spin gap by Oogarah *et. al.*.[16] CF splitting was also widely studied in $Nd_2CuO_4$[10], $NdMnO_3$[11], $R$FeAsO$_{1-x}$F$_x$ ($R$ = Pr, Nd)[17], $YbMnO_3$[18], $HoMn_2O_5$[19], and $DyMnO_3$[12]. It is widely believed that magnetic interactions play critical roles in the CF splitting. However, underpinning the microscopic mechanism underlying unusual CF behaviours in particular rare-earth CMOs is still a challenging subject deserving of further profound research.

Rare-earth orthoferrites $RFeO_3$ (where R represents a rare-earth element) exhibit fascinating magnetism on the two magnetic sublattices of $Fe^{3+}$ and $R^{3+}$.[4,20] Recent discoveries of novel properties in rare-earth orthoferrites,[21-24] including multiferroicity[21,24] and ultrafast laser responses,[25] rekindled research interests in this group of materials. $RFeO_3$ compounds share a perovskite crystal structure[4] with similar orthorhombic distortions, described in the space group *Pbnm*.[3] The $Fe^{3+}$ sublattice forms a canted long-range antiferromagnetic (AFM) magnetic structure at $T_{N(Fe)}$ between 600K and 740K.[4] Most $RFeO_3$ undergoes a spin-reorientation transition at a lower temperature $T_{SR}$ or in a temperature regime between $T_l$ and $T_u$. In some $RFeO_3$, $R^{3+}$ magnetic moments order at significantly lower temperatures ($T_{N(R)}$), near the liquid-helium temperature.[3] It is generally believed that the magnetic behaviour of $RFeO_3$ is a combined result of the interactions of $Fe^{3+}$-$Fe^{3+}$, $R^{3+}$-$Fe^{3+}$, and $R^{3+}$-$R^{3+}$. Single-ion anisotropy and weak antisymmetric exchange interactions play additional roles. [4,26]

In $RFeO_3$, CF excitations were widely observed and investigated. The outermost 4*f* electrons of $R^{3+}$ are split into multiplets due to the spin-orbital coupling. These multiplets could be further split in energy due to the CF introduced by the surround ligands. $Er^{3+}$ in $ErFeO_3$ has an electron configuration of $4f^{11}$ with the lowest multiplet $^4I_{15/2}$, which splits into several doublets and quartets due to the CF effect depending on the local symmetry.[27] Despite the fact that CF excitations in $ErFeO_3$ have previously been described, no thorough investigation has yet been conducted. Recently, Zic et al. discovered that the exchange interaction between the $Er^{3+}$ and $Fe^{3+}$ sublattices caused a slightly dispersive CF excitation at low temperature.[28] Here we use the inelastic neutron-scattering technique to examine the energy shifts of CF excitations in $ErFeO_3$ in the absence and presence of external magnetic fields.

In this study, a variety of external magnetic fields and a wide temperature range were used to characterise the low-energy CF excitation of $Er^{3+}$. By employing Steven's operator equivalents to fit the CF parameters, a CF model was presented. Utilizing this model, the dependences of the CF spectra on temperature, internal field, and external field were simulated and calculated, giving rise to a profound knowledge of the process behind the observed energy shifts in the CF spectra of $Er^{3+}$ in $ErFeO_3$.

## II. EXPERIMENTAL DETAILS

The $ErFeO_3$ single-crystal sample in this study was grown by using the optical floating-zone furnace (FZ-T-10000-H-VI-P-SH, Crystal Systems Corp.) in the Department of Physics at Shanghai University. We performed the inelastic neutron-scattering experiments on the cold-neutron triple-axis spectrometer Sika[29], the thermal-neutron triple-axis spectrometer Taipan[30], and the time-of-flight (TOF) spectrometer Pelican[31] at the Australian Centre for Neutron Scattering (ACNS), Australian Nuclear Science and Technology Organisation (ANSTO). On Sika, a constant-$E_f$ mode with $E_f = 5$ meV and 60'-60'-60'-60' collimation was configured for the experiment. A cooled Be-filter was used to remove the second-order contamination in the scattered neutron beam. On Taipan, we used full open collimation and the constant-$E_f$ mode ($E_f = 14.87$ meV) as the configuration. A 4cm-thick PG (002) filter was placed in front of the pre-analyser collimator to suppress the high-order wavelength contamination. The sample was cooled to the desired temperature using a He-flow cryostat, which was controlled by a Lakeshore 340 temperature controller. The data from both Sika and Taipan were fitted by convoluting with the instrumental resolutions of the configurations mentioned above. The software Octave[32] and the package Reslib3.4[33] were used for the data fitting. The Pelican experiment was carried out with an incident neutron beam of wavelength = 3.65Å. A closed-cycle refrigerator was used to maintain the sample temperature. The Pelican data were treated by using the neutron-data-visualization software LAMP.[34] The CF excitation was fitted by a least-square-minimization python code based on the neutron scattering data analysis package Mantid[35] and the CF calculation package PyCrsytalField[36].

## III. RESULTS AND DISCUSSION

### A. Temperature dependence of low-energy CF excitation

Inelastic neutron scattering experiments were carried out on Sika, and the results are displayed in Fig. S1 in the Supplemental Materials. An excitation below 1 meV was observed in $ErFeO_3$ on Sika (Fig. S1), which shows no apparent dispersion. Thus, it could be safely attributed to a CF excitation from $Er^{3+}$. This peak shows a strong temperature dependence at low temperatures, especially, upon cooling from just above the magnetic ordering temperature ($T_{N(Er)}$) of $Er^{3+}$.

To precisely determine the temperature dependence of this lowest-energy excitation peak, we conducted the inelastic neutron scattering measurement on Pelican from 1.7 K to 10 K with 0.1 K per step. An intensity contour map by combining all these temperature scans is shown in Fig. 1, and the corresponding line plot is shown in Fig. 2(a). The result at each temperature is the summation over the full accessible $Q$ range in the experiment. The strong peak at zero energy transfer corresponds to the elastic scattering from the sample. The linewidth of this peak was considered as the instrumental resolution for this configuration. A strong excitation peak was observed around 0.75 meV at 1.5 K. This peak shows no dispersion, which is verified by inspecting energy slices at the different $Q$ cuts from the three-dimensional $S(\boldsymbol{Q}, \omega)$ data generated from the Pelican experiment. This result is also consistent with the experimental data collected on Sika, as presented in Fig. S1 in the Supplemental Materials.

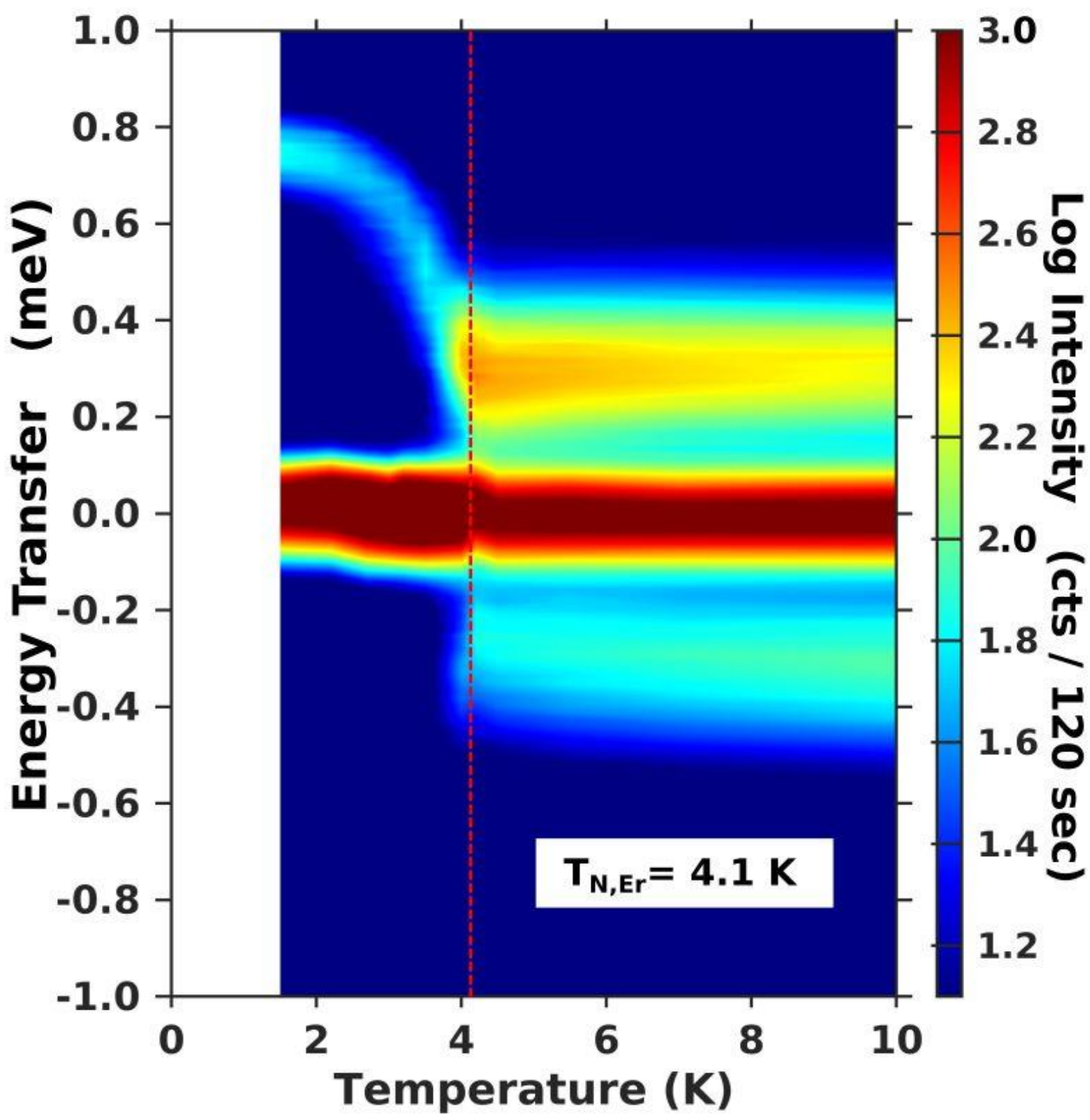

FIG. 1. (Colour online) False-colour contour map of the intensities of the inelastic neutron scattering data of $ErFeO_3$ measured on the time-of-flight spectrometer Pelican at ANSTO at selected temperatures in the vicinity of $T_{N(Er)}$ = 4.2 K. $T_{N(Er)}$ is marked by the red vertical dashed line. The excitation visible at ± 0.35 meV at 10 K is attributed to a CF excitation. Its excitation energy significantly increases at temperature below $T_{N(Er)}$ while its integrated intensity and linewidth decrease. The red bar at zero energy transfer is the central peak from the elastic scattering of the sample. The intensity, which indicated by the colour bar, is based on a logarithmic scale.

When heating the sample gradually from 1.5 K to 4.2K, namely, $T_{N(Er)}$, the excitation energy decreases gradually to ~0.35 meV. On further heating, the excitation energy of this peak does not show strong temperature dependence up to 10K. The fitted peak positions at different temperatures are plotted in Fig. 2(b). In contrast to the single excitation peak observed below $T_{N(Er)}$, the CF excitations were observed on both the energy-loss and energy-gain sides at temperatures above $T_{N(Er)}$, which correspond to the Stokes and anti-Stokes excitations, respectively. This means that more spins occupied in the exited state at higher temperature could transfer energy to neutrons and jump back to the low-energy state. The excitation energy demonstrates a similar temperature dependence as the magnetic moment of $Er^{3+}$ obtained from the previous neutron powder diffraction experiment.[3] Such a similarity undoubtedly indicates that the CF excitation is strongly correlated with the long range magnetic ordering of $Er^{3+}$ spins. Such a large energy shift driven by magnetic order is quite unique and worthwhile investigating its underlying mechanism.

Fig. 2 (c) and (d) shows the temperature dependences of the linewidth and intensity according to the fittings, respectively. Both decreases significantly below $T_{N(Er)}$. The narrowing linewidth could be attributed to the suppressed thermal fluctuation of the $Er^{3+}$ magnetic moments below $T_{N(Er)}$, as pointed out by Berker et al.[14] The abrupt decrease of the intensity at the transition is obviously due to the ordering of $Er^{3+}$ spins. In the temperature regime of $Er^{3+}$ magnetic ordering, the peak intensity gradually recovers upon further cooling. Similar temperature dependence was widely reported in other rare-earth systems[37,38]

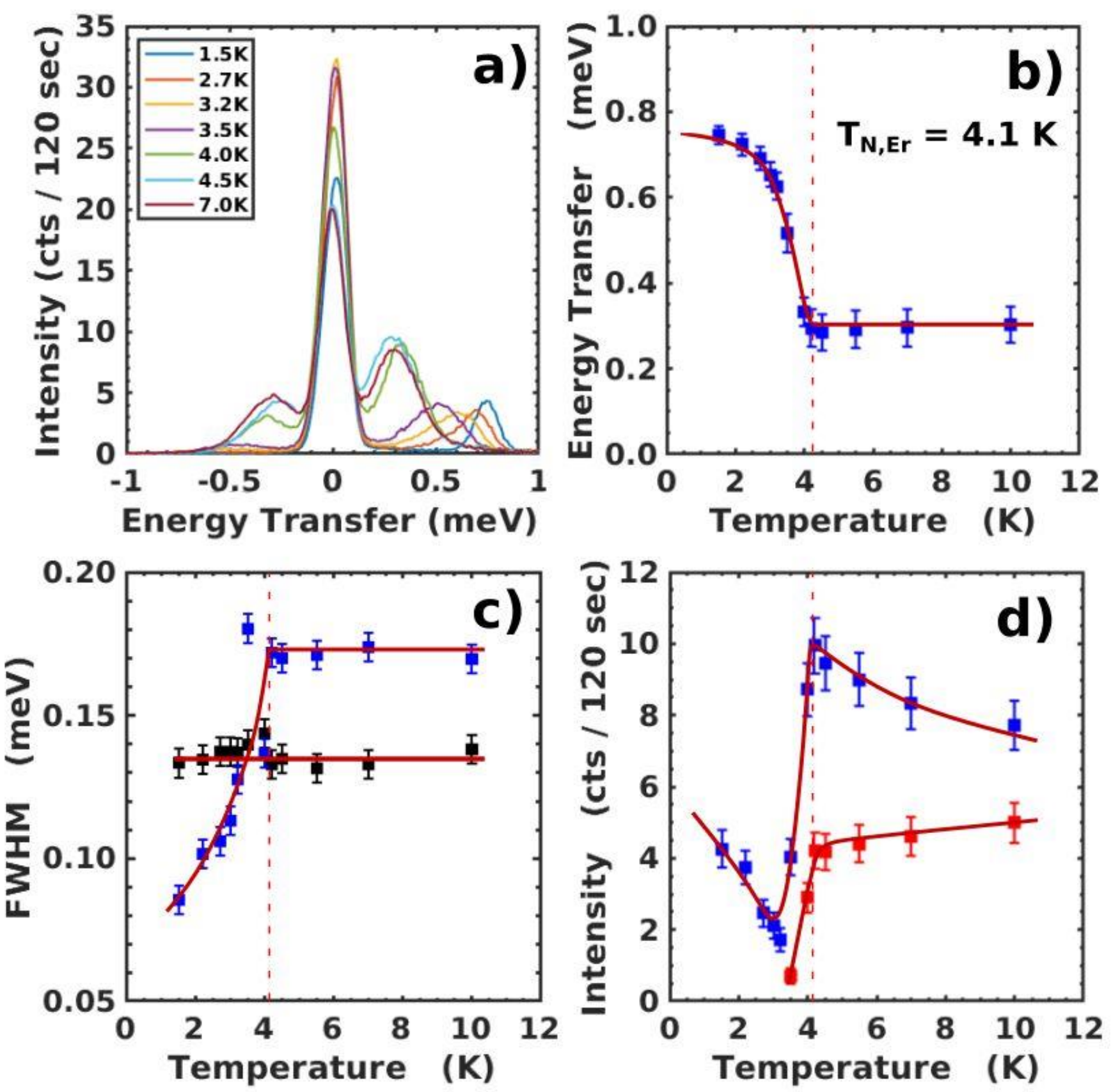

FIG. 2. (Colour online) Temperature dependence of the low-energy CF excitation of $ErFeO_3$. (a) Excitation spectra at the selected temperatures, (b) Temperature dependence of the CF excitation energy determined by fitting to the experimental data; (c) Temperature dependence of the CF (blue symbols) and central peak (black symbols) linewidths; (d) Temperature dependence of the Stokes (blue symbols) and anti-Stokes (red symbols) peak intensities. The red solid lines in (b), (c) and (d) serve as guide to eyes while the vertical red dash lines indicate $T_{N(Er)}$ at ~ 4.2K.

The significant energy shift of the CF excitation observed above strongly correlates to the ordering of $Er^{3+}$ moments in $ErFeO_3$. Such a strong correlation must play roles in other CF excitations as well. To understand the underlying mechanism of this correlation, one needs to consider the full picture of the CF excitation model of $ErFeO_3$.

### B. CF Model of $Er^{3+}$ in $ErFeO_3$

To fully understand the CF energy shift discussed above, we carried out further measurements and performed CF modelling. The CF excitations from 0 to 10 meV were measured in $ErFeO_3$ on Taipan at different temperatures from 1.5 K to 150 K. The collected data and the fittings to the data are plotted in Fig. S4. The fitted peak positions and intensities are presented in Fig. S5(a) and (b), respectively. These figures clearly exhibit the energy shift of another CF excitation peak at ~5.75 meV. At elevated temperatures from 10K to 150K, the measurements show another nondispersive excitation peak at ~7.5 meV, the intensity of which increases upon heating. In contrast, the peak intensity at ~5.75 meV decreases when the temperature rises. Therefore, we speculate that the peak at ~7.5 meV corresponds to an excitation from one excited CF state to another excited CF state in $ErFeO_3$. A sound CF model is indispensable to validate this speculation. Following this idea, we carried out a wide energy scan from 2.5 meV to 30 meV at 1.5 K on Taipan to catch more CF excitation peaks. Fig. 3 shows this wide-range scan at an off-centre Q position, which reveals four intense peaks at ~5.75 meV, ~13.5 meV, ~20.5 meV, and ~25 meV (see Fig. S6). These observed CF excitation levels are consistent with the previously published fluorescence results.[39] In the following, we attempt to build a CF model to interpret the CF excitations of $Er^{3+}$ in $ErFeO_3$

CF excitations of 4*f* ions can be normally considered as a perturbation of its lowest-energy multiplet because the spin–orbital coupling is much stronger than the CF effect in 4*f* ions. $Er^{3+}$ has a $^4I_{15/2}$ ground state and the first excited state is $^4I_{13/2}$. Since $Er^{3+}$ is a Kramers ion, the $Er^{3+}$ ground state multiplet split into two doublets ($\Gamma_6$ and $\Gamma_7$) and three quartets ($\Gamma_8$) in a cubic CF symmetry.[40] Ammerlaan and de Maat-Gersdorf studied the $Er^{3+}$ CF splitting in different local symmetric environments, such as cubic, trigonal, tetragonal and orthorhombic fields.[41] Their numerical results demonstrate that the orthorhombic CF of $Er^{3+}$ can be described as a perturbation of the tetragonal CF, which is denoted by an extra term in its Hamiltonian. The ground state multiplet in orthorhombic symmetry splits into eight Kramers doublets except for some accidental degeneracies. Similarly, a small distortion of lower symmetry can be considered as a perturbation of the orthorhombic CF. This method was successfully used to describe the CF excitation of $Er^{3+}$ on the interstitial sites.[40] In $ErFeO_3$, $Er^{3+}$ occupies the 4e Wyckoff position and has the point group of Cs. Generally, the Hamiltonian of CF reads as follows:

$$H_{CF} = \sum_{k,q} B_k^q O_k^q(J) = \sum_{k,q} A_k^q \langle r^n \rangle \theta_k O_k^q(J) \quad (1)$$

where $B_k^q$ are the CF parameters, and $O_k^q$ are Stevens operator equivalents which are written in the powers of the angular moment operators, $J^+$, $J^-$ and $J_z$. $A_k^q$ are the CF coefficients, and $\langle r^n \rangle$ is a radial distribution function of the 4*f* electrons with $k$=2, 4, 6. $\theta_k$ are operator equivalent factors for $k$ = 2, 4, and 6. For the CF of $Er^{3+}$ on $C_s$ sites, Eq.(1) takes the reduced CF parameters listed in TABLE I[40].

The $B_k^q$ values were estimated as the initial parameters for the fitting by using the point-charge model,[42,43] Using a python non-linear least-squares minimization program developed on the basis of Mantid,[36] we fitted the experimental data to the CF model above and obtained the optimized CF parameters as shown in TABLE II. The corresponding CF levels were calculated and listed in TABLE III. The obtained model was checked and confirmed by another CF calculation package PyCrystalField, indicating the consistent results from both these packages.

TABLE I Reduced CF parameters for $Er^{3+}$ in the $C_s$ point group in $ErFeO_3$

| *k* | *q*=0 | *q*=2 | *q*=4 | *q*=6 |
|---|---|---|---|---|
| 2 | $B_2^0$ | $B_2^{\pm 2}$ | | |
| 4 | $B_4^0$ | $B_4^{\pm 2}$ | $B_4^{\pm 4}$ | |
| 6 | $B_6^0$ | $B_6^{\pm 2}$ | $B_6^{\pm 4}$ | $B_6^{\pm 6}$ |

For comparison, Fig. 3 plots the background-subtracted experimental data (the blue-symbol curve, see Fig. S3 for more details about this data), the fitted curve (the red curve), and the calculated spectrum according to the CF model (the green curve) together. The fitted curve matches the experimental data quite well. The calculated spectrum presents not only the peaks at 5.78, 13.42, 20.56, and 25.07 meV, but also the peak at the zero-energy transfer corresponding to the ground CF level. The peak positions and intensities of all four fitted peaks highly agree with the observed peaks. However, this model did not predict the peak below 1 meV and the peak near 18.5 meV, both of which were observed in the experiment. The reason for these two peaks showing up in the experiment will be discussed in detail later.

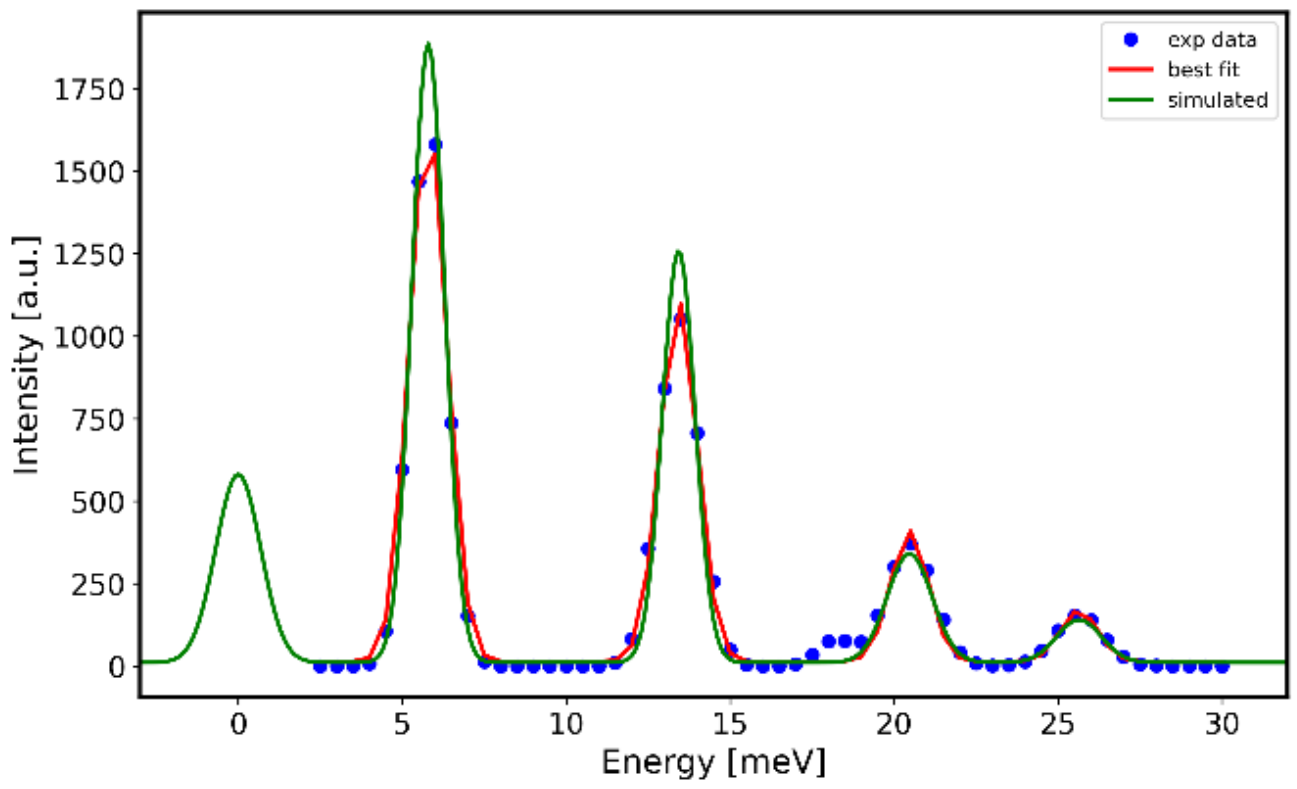

FIG. 3. The CF excitations measured in $ErFeO_3$ at 1.5K, which are fitted to the CF model described in the article. The blue dotted curve shows the experimental data after subtracting the background. The red curve is the fitted result to the experimental data and the green curve is the simulated CF spectrum of $Er^{3+}$.

In Fig. S2 (a), the experimental data at enhanced temperatures also show an excitation peak at ~7.5 meV, which grows upon heating. This provides another fact to check the model obtained above. We attributed this peak to the excitation from the excited level at ~ 5.75 meV to the other excited level at ~13.5 meV. The CF excitation spectra of $Er^{3+}$ in $ErFeO_3$ were calculated at a series of temperatures using the CF model above. The spectra were plotted in Fig. S7, in which a new peak clearly shows up on the high energy side of the peak at ~5.75 meV at elevated temperatures. It matches the observed peak at this temperature so well. Such a consistency between the model and the experiment strongly supports the validity of the CF model we proposed here. Based on this CF model, we plot the energy scheme diagram of the $Er^{3+}$ CF levels in Fig. 4. It shows the zero-energy ground level and all other excited levels. The excitation at ~7.5 meV was marked as a green line indicating that this excitation is from the ground level.

TABLE I Fitted CF parameters of the CF model. The unit for these parameters is meV.

| $B_2^0$ | $B_2^2$ | $B_2^{-2}$ | $B_6^{-4}$ | $B_6^{-6}$ |
|---|---|---|---|---|
| -0.3313 | -0.526 | 8.275e-07 | -9.931e-06 | 4.394e-06 |
| $B_4^0$ | $B_4^2$ | $B_4^4$ | $B_4^{-2}$ | $B_4^{-4}$ |
| 3.21e-04 | 7.31(8)e-03 | 4.15e-03 | -1.876e-04 | -1.821e-04 |
| $B_6^0$ | $B_6^2$ | $B_6^4$ | $B_6^6$ | $B_6^{-2}$ |
| 2.87e-06 | -1.074e-04 | -1.501e-04 | 1.498e-04 | 9.563e-06 |

TABLE II Calculated CF energies and intensities of eight $Er^{3+}$ Kramers doublets in $ErFeO_3$

| | 1 | 2 | 3 | 4 | 5 | 6 | 7 | 8 |
|---|---|---|---|---|---|---|---|---|
| CF level (meV) | 0.0 | 5.79 | 13.43 | 20.47 | 25.64 | 49.38 | 76.00 | 131.31 |
| Intensity (a.u.) | 758.94 | 1817.53 | 1209.69 | 438.49 | 172.24 | 20.39 | 13.12 | 1.82 |

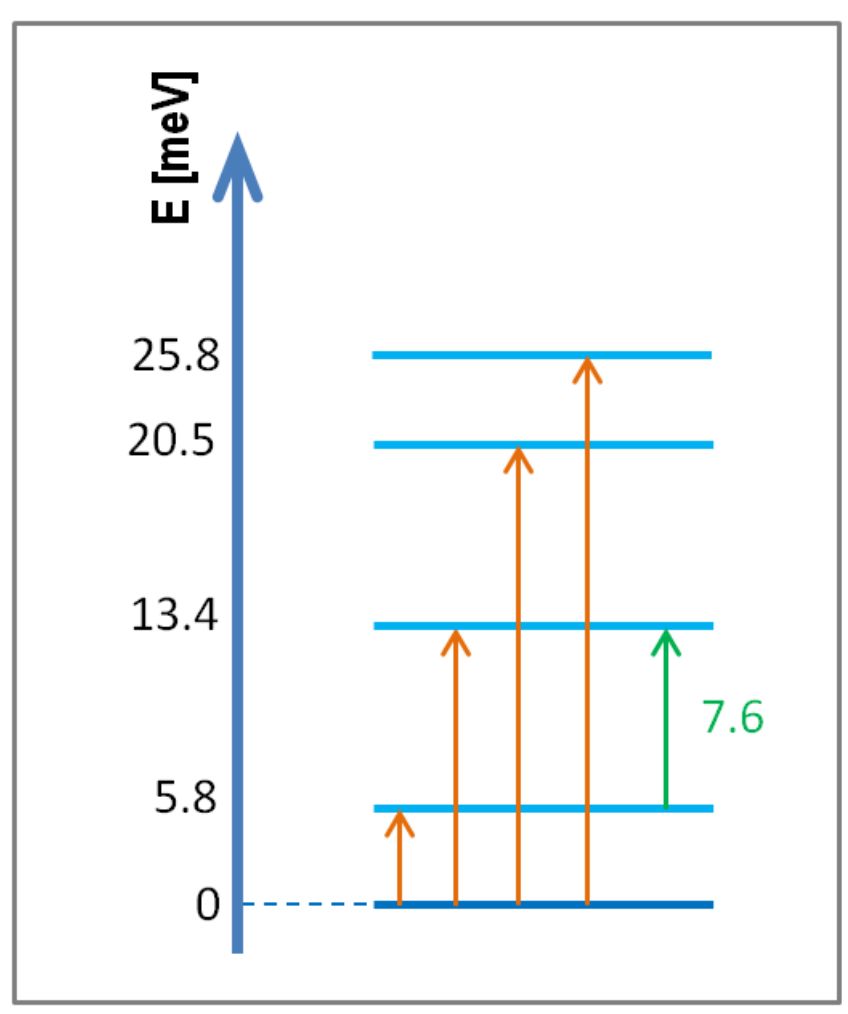

FIG. 4. CF energy level diagram of $Er^{3+}$ in $ErFeO_3$ in the energy range from 0 to 30 meV according to TABLE III. Each horizontal lines in this diagram corresponds to one of the eight degenerated CF doublets of $Er^{3+}$. The solid dark line at the bottom denotes the ground CF level at 0 meV. The other blue lines are the excited CF levels observed in the experiment. The orange arrows indicate the excitations from the ground level while the green arrow shows the excitation from the second CF level to the third level, which was observed only at the elevated temperatures. The CF energy levels higher than 30 meV in TABLE III are not shown in this diagram.

The CF model produces the eigenfunctions for all the CF levels of $Er^{3+}$ in $ErFeO_3$. The ground level with the three largest eigenvector components reads $A = 0.5392\left|-\frac{15}{2}\right\rangle - 0.4169\left|-\frac{7}{2}\right\rangle + 0.1376\left|-\frac{9}{2}\right\rangle$. Since $ErFeO_3$ has a canted antiferromagnetic ground state with $Er^{3+}$ spins ordering at base temperature, the internal field of $ErFeO_3$ may play roles in the CF excitation as well. The CF excitation spectra were simulated at a variety of temperatures and plotted in Fig. 5. The simulation clearly demonstrates the splitting and shifting of the ground CF level in the fields along the $x$ and $y$ axes. The first excited state (~ 5.75 meV) splits and shifts when the internal fields are induced along the $z$ axis. The peak at ~ 20.5 meV undergoes similar splitting or shifting in fields along all three different directions. Considering the two unexplained peaks observed at 0.35 meV and 18.5 meV in the experiment, we could attribute these two peaks to the effect of the internal field in $ErFeO_3$. Fig. 5 demonstrates that an internal field ~ 2.5 T along the $y$ axis (namely, the $b$ axis) can shift the ground CF level peak from 0 to ~ 0.35 meV and create a shoulder at ~18.5 meV on the low-energy side of the peak at ~ 20.5 meV. Surprisingly, such an internal field along the $b$ axis do not make much impact to the other two peaks at ~ 5.75 meV at 13.5 meV. Such a behaviour is highly consistent with the experimental data, strongly supporting the scenario that the two peaks below 1 meV and near 18.5 meV were due to the internal field effect. As we know from in our previous study, $ErFeO_3$ has two magnetic phases corresponding to $Fe^{3+}$ and $Er^{3+}$ sublattices at the base temperature. Since $Er^{3+}$ spins have been polarized by the magnetically ordered $Fe^{3+}$sublattice through the $Fe^{3+}$-O-$Er^{3+}$ exchange interaction at the temperature range well above $T_{N(Er)}$,[44] the $Fe^{3+}$ magnetic order indirectly induces the internal fields on the $Er^{3+}$ sites, too. Therefore, both $Er^{3+}$ and $Fe^{3+}$ magnetic orders could produce internal magnetic fields to shift or split the CF excitations. Above $T_{N(Er)}$, the internal field induced by the $Fe^{3+}$ spin order drives the ground level from 0 to ~ 0.35 meV while below $T_{N(Er)}$, the internal field from the $Er^{3+}$ spin order plays an additional role to push this peak further to ~ 0.75 meV. From the simulation, this scenario seems highly reasonable for our observation.

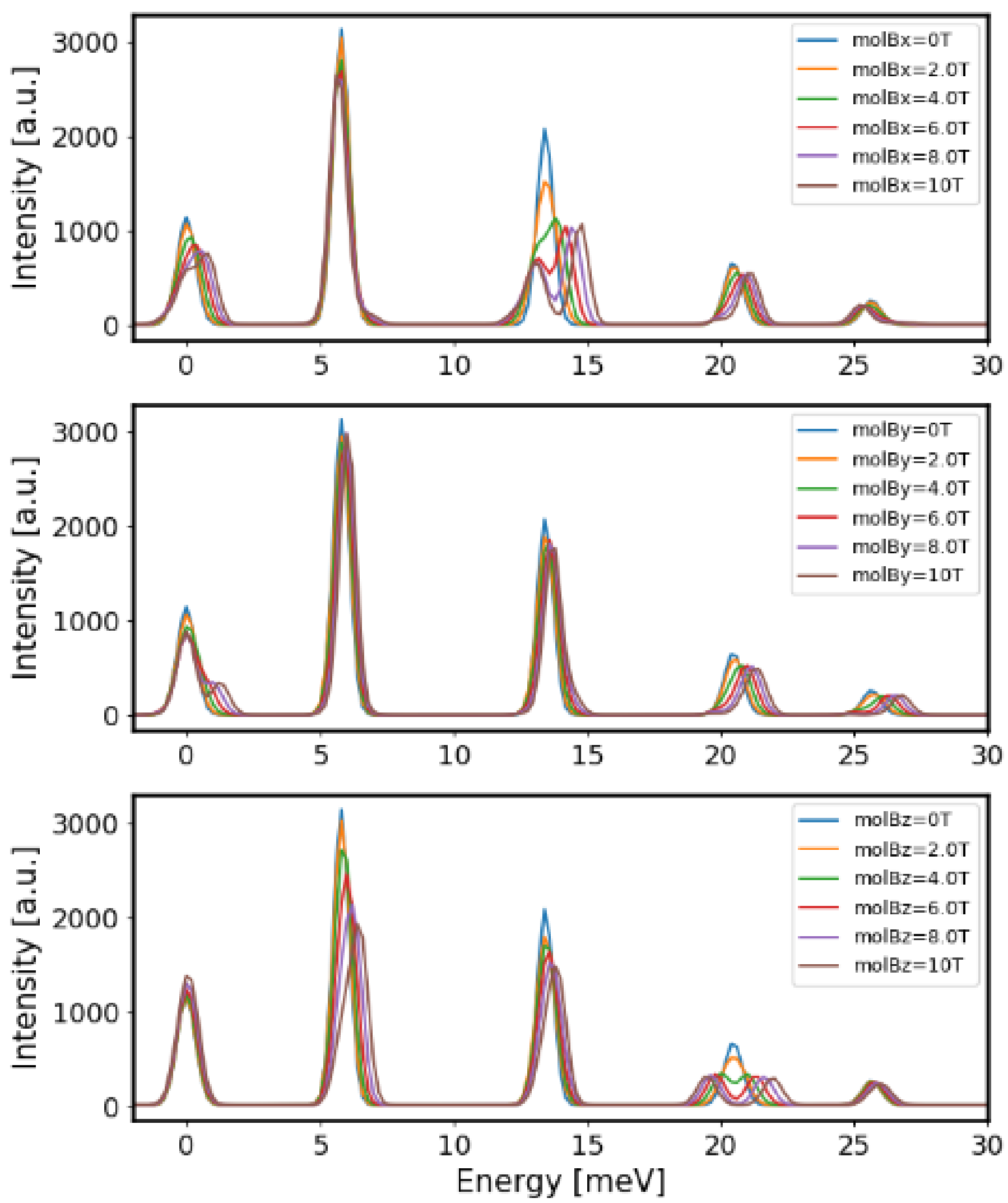

FIG. 5. Simulated internal molecular magnetic field effects on the $Er^{3+}$ CF energy levels with the internal fields along the (a) *x*, (b) *y,* and (c) *z* directions.

**C. Effect of an external magnetic field on the CF excitation of $ErFeO_3$**

As one step further, we conducted inelastic neutron scattering experiments under external magnetic fields on Sika. The $ErFeO_3$ single crystal was mounted with the *ac* plane as the scattering plane. Vertical magnetic fields from 0 up to 10 T were applied step by step for the measurements. The magnetic field dependencies of the lowest-energy excitation were measured at both 1.5K and 10K, below and above $T_{N(Er)}$, respectively. The results are summarized in Fig. 6. Fig. 6 (a) and (c) show the magnetic-field dependences of the CF excitation energy in the high-field region up to 10 T at 1.5 K and 10 K, respectively. Fig. 6 (b) shows the low-field dependence of the CF excitation energy in the low external-magnetic-field range with more details.

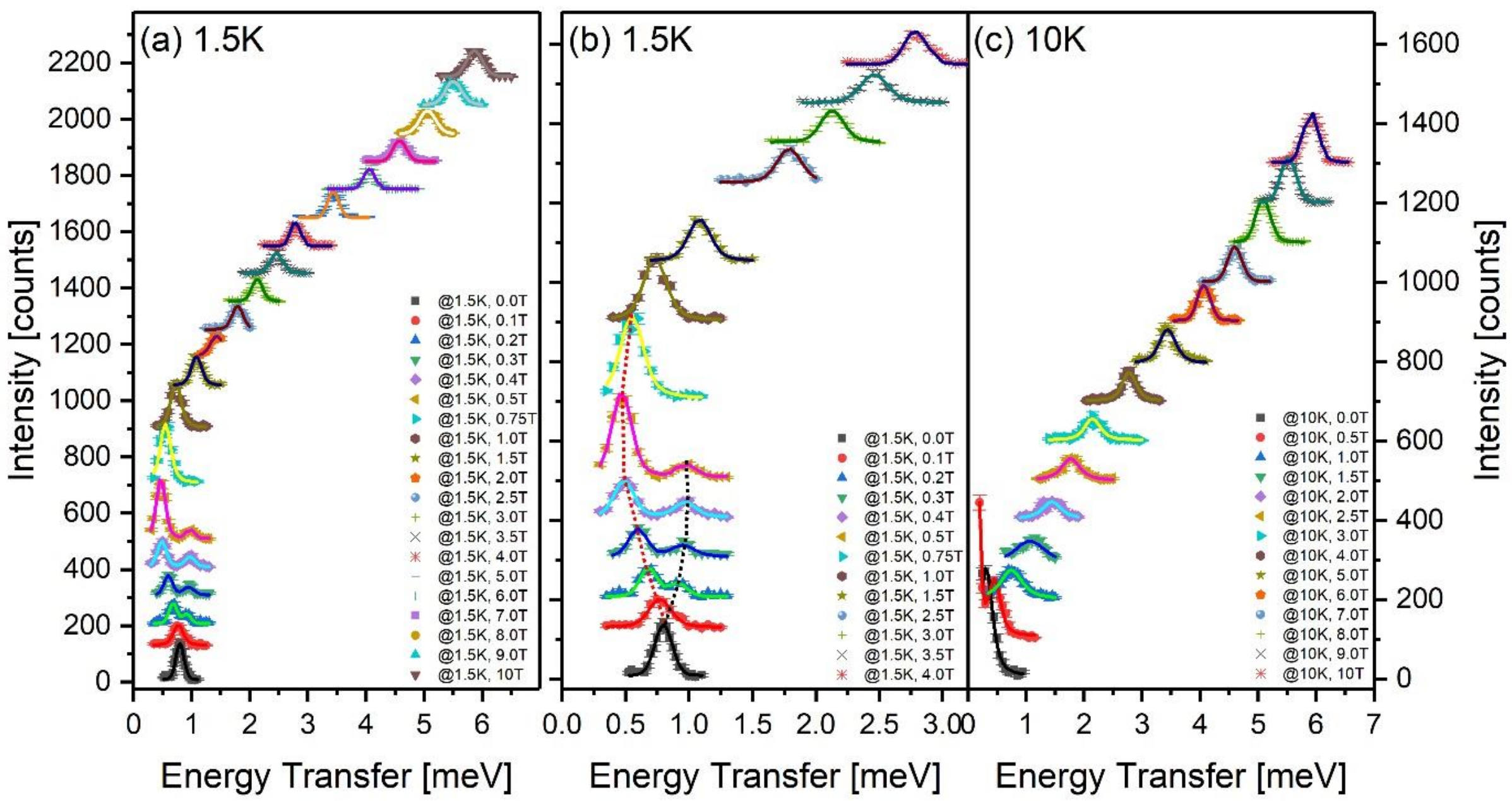

FIG. 6. (colour online) Magnetic-field dependences of the low-energy CF excitation of $ErFeO_3$ were determined by conducting energy scans at $Q$(1, 0, 0.5) on Sika at (a), (b) 1.5K and (c) 10K, respectively. The magnetic fields were applied along the crystallographic $b$ axis. The symbol curves in (a), (b) and (c) are the experimental data, which are fitted to the damped harmonic oscillator model by convoluting with the instrumental resolution. The solid colour lines show the fitted curves. (b) shows the magnified low-field zone of (a) with the same intensity scale as (c). The dotted lines in (b) are guides to eyes. The scans in each of (a), (b) and (c) are offset along the vertical axis to avoid the individual scans overlapping each other.

At 1.5K, $Er^{3+}$ magnetic moments order into the $C_z$ antiferromagnetic structure. The single peak at zero field shows a distinct splitting into two branches when applying a low magnetic field (< 1T). Upon increasing the external fields within this range, the energy of the upper branch gradually increases while its intensity decreases correspondingly, and the peak eventually vanishes at ~ 0.6 T. The energy of the lower branch first decreases from 0.75 meV to an energy level of about 0.5 meV when increasing the external magnetic fields from 0 to 0.5 T. Above 0.5T, the CF excitation energy of this branch starts to increase almost linearly with the applied external field. The data in Fig. 6(a) were fitted by convoluting with the instrumental resolution. The fitted peak positions are plotted in Fig. 7(a), in which the blue-symbol curve corresponds to the upper branch and the red one represents the lower branch. This figure clearly exhibits the linear relationship of the lower-branch energy with the applied external field in the range from 1T to 6T. The slope of this linear relationship slightly dropped at the higher magnetic field range from 6T to 10T. At the external field of 10T, the CF excitation peak was observed at ~5.8 meV.

The results from the paramagnetic state of $Er^{3+}$ at 10K were fitted and plotted in Fig. 6(b) by using the same convolution method and the fitted results were plotted in Fig.7(b). Major differences between the 10 K and 1.5 K data appear at magnetic fields below 1T (see Fig. 6 and 7). At 10K, no splitting of the CF excitation peak was observed under external fields, which suggests that the ordering of $Er^{3+}$ magnetic moments caused the splitting of the lowest CF peak under external magnetic fields. Upon applying magnetic fields, the peak position shows a very similar tendency against the fields as the trend observed at 1.5 K in this range. Both have a linear relationship with a slope about 0.65 meV/T from 1 T to 6 T, then the slope slightly drops in the range from 6 T to 10 T. A similar linear relationship was previously reported for the magnetization under external magnetic fields (||*b* axis) from 0 to 6T. The slope slightly decreases in the range from 6 T to higher fields, too.[45] According to this study, the $Er^{3+}$ magnetic moments start to cant along the external field when the field is larger than 6T at ~1.5 K. The rotation of the $Er^{3+}$ magnetic moments could be the reason why the Zeeman splitting energy shows a nonlinear field dependence at fields higher than 6T. The linear and nonlinear field dependencies were observed at both 1.5 K and 10 K. The similarity between the results at 1.5 K and 10 K at high fields may indicate that the ordering effect of $Er^{3+}$ on the CF excitation can be overcome by applying an external field of ~ 1 T.

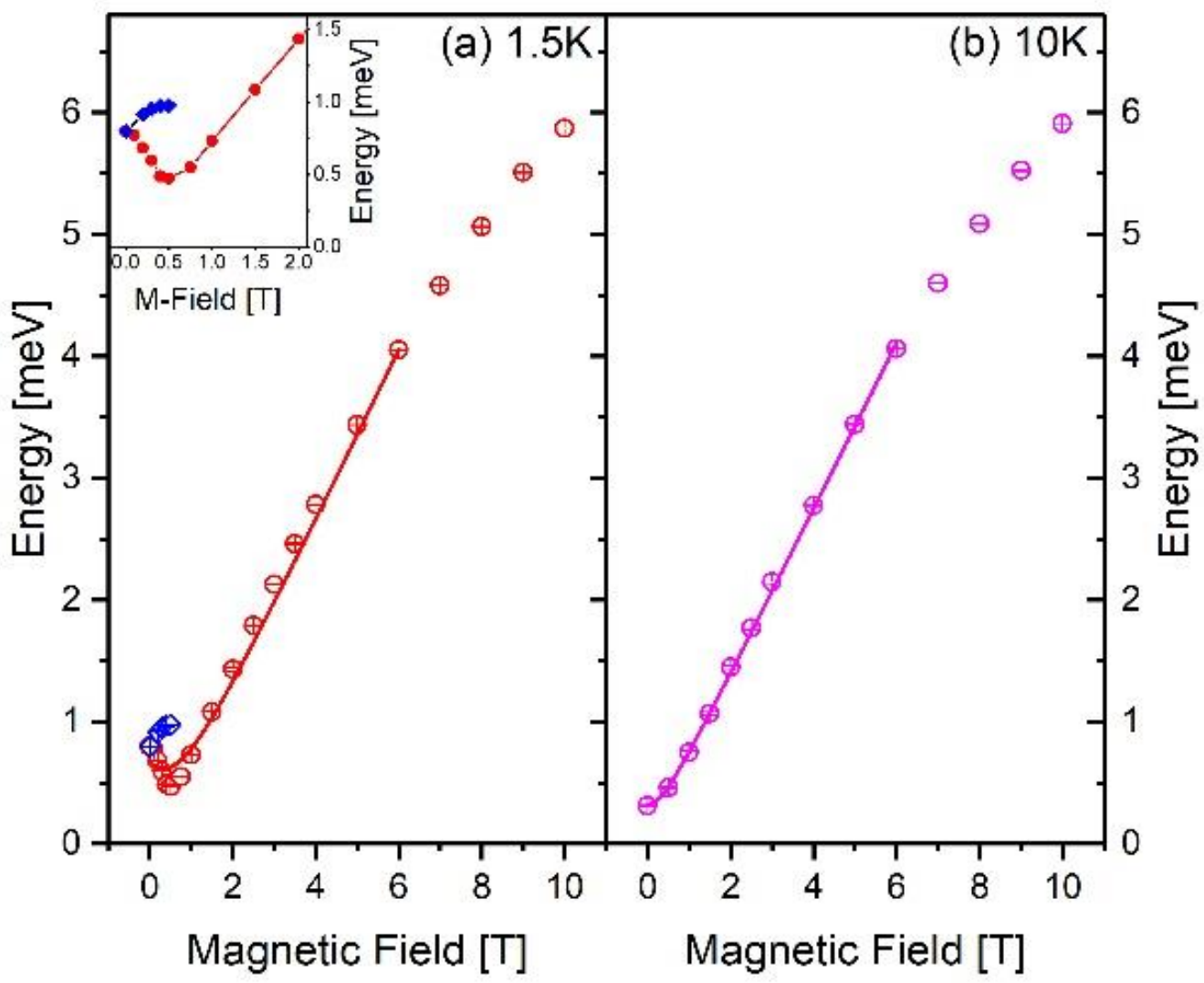

FIG. 7. (Colour online) Low-lying CF excitation energy as a function of externally applied magnetic fields along the crystallographic *b* direction at (a) 1.5K and (b) 10K. The symbols in (a) and (b) shows the fitted CF excitation energy from the experimental data shown in Figures 5(a) and 5(b) at 1.5K and 10K, respectively. The CF excitation splits into two peaks at low magnetic fields at 1.5K, which are denoted by the red and blue symbols in (a). The CF excitation does not split but shifts at 10K as shown by the magenta symbols in (b). The solid lines in (a) and (b) are fitted to the model described in the text. The inset in (a) shows the zoomed-in view of the low -field and low-energy region of (a). The upper branch of the split peak is hardly discerned at the external fields above 0.5T.

Differently from the three quartets and two doublets in cubic symmetry, the $Er^{3+}$ ground state splits into eight Kramers doublets in $ErFeO_3$. These doublets with Kramers degeneracy can be lifted when applying an external magnetic field, or due to the internal fields induced by the long-range magnetic ordering. Such splitting of the degenerated doublets has been reported for many rare-earth compounds.[46,47] For example, Iwasa *et al.* reported the splitting of the non-Kramers doublet of $Pr^{3+}$ in $PrT_2Zn_{20}$ (T = Ir, Rh, and Ru) by using inelastic neutron scattering.[48] The splitting of the ground-state Kramers doublet of $Sm^{3+}$ was observed at 2K in $SmMnO_3$ by Nekvasil et al.[49] The ground-state Kramers-doublet splitting in $DyMnO_3$ was reported by Jandl et al. [12] with their infrared transmission measurements.

The ground level CF doublet of $Er^{3+}$ is sensitive to the internal fields in $ErFeO_3$ at low temperatures. When the spins of rare-earth ions form a long-range ordering, the internal field builds up and shifts the CF energy. Considering both the effects of the internal and external magnetic fields, the splitting energy 2Δ of the doublet can be determined by the following formula:[46]

$$\Delta^2 = \boldsymbol{\Delta}_{cf}^2 + \left[\boldsymbol{g}_{eff}\boldsymbol{\mu}_B(\boldsymbol{H}_{ex} + \boldsymbol{H}_{in})\right]^2 \qquad (3)$$

where $\Delta_{cf}$ is the CF excitation energy, $g_{\mathrm{eff}}$ is the component of the $g$ tensor along the field $H$. $\mu_B$ is the Bohr magneton, $H_{\mathrm{ex}}$ is the applied external magnetic field, and $H_{\mathrm{in}}$ is the component of the effective internal magnetic field along $H_{\mathrm{ex}}$. Since the $Er^{3+}$ magnetic moments stay in an antiferromagnetic order at 1.5K, there are two effects to the local CF excitations. A half of the $Er^{3+}$ moments generate an internal field adding up to the external magnetic field while another half create an internal field opposite to the external field. Thus, two sets of local fields are created and result in the splitting effect in the magnetic fields lower than 0.5 T. When the external field increases to a level much larger than the internal field, no splitting will be observed. It is worthwhile to mention that the $g_{\mathrm{eff}}$ factor, which is the component along the external magnetic field, generally varies for different CF levels, different magnetic fields, and different temperatures, too. Fitting the experimental data in Fig. 7 to Eq. (3), we obtained the fitting parameters shown in TABLE IV:

TABLE IV. Fitted exchange energy, internal field and $g_{\mathrm{eff}}$ factor from Eq. (3) at 1.5K and 10K

| T [K] | $\Delta_{cf}$ (meV) | $H_{\mathrm{in}}$ (meV) | $g_{\mathrm{eff}}$ |
|---|---|---|---|
| 10 | 0.32 (3) | 0 | 11.79(5) |
| 1.5 | 0.62 (5) | 0.33 (13) | 12.20 (28) |

The second row of TABLE IV shows the fitted parameters from the data at 10K. We set the parameter $\boldsymbol{H}_{\mathrm{in}}$ to zero and the fitting gives $\Delta_{cf} = 0.32 \pm 0.025$ meV and the $g_{eff}$ component = 11.79. The $\Delta_{cf}$ value is consistent with the observed energy gap at 10K, which is ~ 0.35 meV, indicating that a uniform internal field of the similar level exists in $ErFeO_3$ at this temperature. The $g_{eff}$ factor is substantially larger than the theoretical values of 5.6 ($\Gamma_6$) or 6.8 ($\Gamma_7$) proposed in reference[41] , which indicates the strong spin-orbital coupling of $Er^{3+}$ ions in this magnet.

With the antiferromagnetic order of the $Er^{3+}$ spins at 1.5K, the CF level has the impact from the external field and the internal field from $Er^{3+}$. Fitting the experimental data at 1.5K to Eq. (3) generates the parameters listed in the third row of TABLE IV. The excitation energy $\Delta_{cf}$ is ~ 0.62 meV, the internal field $H_{\mathrm{in}}$ ~ 0.33 meV and $g_{\mathrm{eff}}$ ~ 12.2. Here, $\Delta_{cf}$ is much larger than its value at 10K. The internal field $H_{\mathrm{in}}$ is comparable to the difference between the two values of $\Delta_{cf}$ at 1.5K and 10K. This gives a rough estimation about the internal field created by the $Er^{3+}$ magnetic order. The $g_{eff}$ values at both 1.5K and 10K are almost the same and much larger than the reported $g_{eff}$ values at 77K and the theoretical values.[39] Similar large $g_{eff}$ values have been reported previously,[41] which is mainly due to the large anisotropy of the $g_{eff}$ factor induced by the spin-orbital coupling.

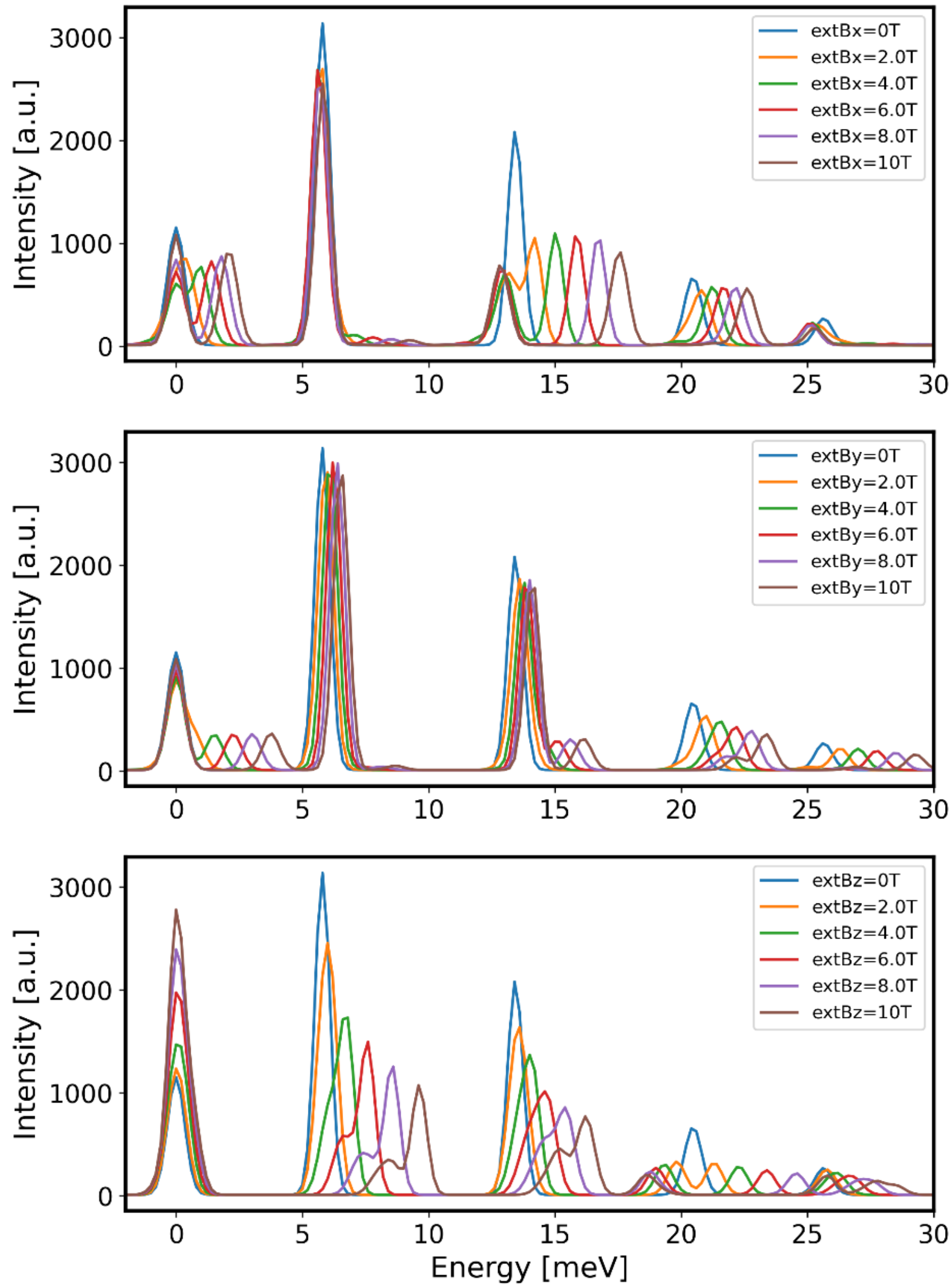

FIG. 8. Simulated external magnetic field effects on the $Er^{3+}$ CF energy levels with the fields applied along the (a) *x*, (b) *y,* and (c) *z* directions.

As known from above, the CF levels of $ErFeO_3$ splits into eight doublets in $ErFeO_3$. The excitation peak below 1 meV at zero magnetic field is shifted to the higher energy by the internal field induced by the ordered $Er^{3+}$ magnetic moments when the sample is cooled below $T_{N(Er)}$. Applying low external fields at 1.5K can split this excitation further in two peaks due its Kramers-doublet nature. The Zeeman splitting below $T_{N(Er)}$ was previously investigated using Mössbauer spectroscopy, resulting in ~0.66 meV at 1.5K and ~0.28 meV at 4.2K, these values are very close to the values we obtained here: 0.62 meV at 1.5K and 0.32 meV at 4.5K.[50] This agreement strongly supports our discussion above.

Similar energy shifts due to the internal field were observed for the other CF levels in $ErFeO_3$. For example, we observed energy shifts with temperature for the CF peak at ~ 5.75 meV, as shown in Fig. S5(a). The big shift step (~ 0.33 meV) from 1.5K to 10K should be ascribed to the internal field change induced by the $Er^{3+}$ ordering. The second step (~ 0.1 meV) from 80K to 100K is due to the internal-field change of the $Fe^{3+}$ sublattice (see Fig. S5(a)) because it undergoes a spin-reorientation in the temperature regime. This indicates that the ordering of $Er^{3+}$ moments and spin reorientation of $Fe^{3+}$ moments strongly affect the CF excitations. It is worthwhile to point out that the effect of the magnetically ordered $Fe^{3+}$ sublattice on the CF excitations of $Er^{3+}$ cannot be excluded in the whole temperature range. However, this effect is smaller than that induced by the $Er^{3+}$ sublattice at low temperature.

## IV. CONCLUSION

Using the inelastic neutron scattering technique, the CF excitations in $ErFeO_3$ were systematically measured at various temperatures and in external fields. When cooling the sample down through the antiferromagnetic phase transition of $Er^{3+}$ spins, a significant energy shift of the lowest-energy CF excitation in $ErFeO_3$ was observed. A CF model is proposed for $Er^{3+}$ in $ErFeO_3$ and explains the detected CF peaks well. According to this model, the generated excitation spectra at various temperatures, different internal magnetic fields and external magnetic fields closely match the experimental data. Analysis of the model and the experimental data reveal that the internal field created by the ordering of $Er^{3+}$ magnetic moments is the driving force for the significant energy shifts of the CF excitation. Fitting the magnetic field dependence data to the Zeeman splitting model, the internal field of the $Er^{3+}$ ordering state was extracted. The significant field dependence of the excitation peak is attributed to the abnormal anisotropy of the $g_{eff}$ factor and sizable $g_{eff}$ component along the *b* axis. This study offers profound a thorough understanding to the impacts of the internal and external fields on the CF excitations in $ErFeO_3$, which may be potentially applied to other TMOs with related excitations.

## ACKNOWLEDGMENTS

Some co-authors would like to thank Drs. Richard, Mole (ANSTO), Rob Robinson and G. Khaliullin (Max Planck Institute for Solid State Research, Stuttgart,) for fruitful discussions. We thank ANSTO for the allocation of neutron beam time on Taipan, Sika, and Pelican (P3169, P3847, P5242, P5295). This work was financially supported by the National Natural Science Foundation of China (NSFC, Nos. 11774217, 12074242, 12074241), the Australian Institute of Nuclear Science and Engineering Ltd (AINSE) and the Australian Research Council (ARC) through the funding of the Discovery Grants DP110105346 and DP160100545.

# Supplemental Material: Giant Shifts of Crystal-field Excitations in $ErFeO_3$ Driven by Internal Magnetic Fields

*Joel O'Brien,[1] Guochu Deng,[2*] Xiaoxuan Ma,[3] Zhenjie Feng,[3] Wei Ren,[3] Shixun Cao,[3†] Dehong Yu,[2] Garry J McIntyre,[2] Clemens Ulrich[1]*

[1] *School of Physics, The University of New South Wales, Kensington, NSW 2052, Australia*

[2] *Australian Centre for Neutron Scattering, Australian Nuclear Science and Technology Organisation, New Illawarra Road, Lucas Heights, NSW 2234, Australia*

[3] *Department of Physics, International Centre of Quantum and Molecular Structures and Materials Genome Institute, Shanghai University, 99 Shangda Road, Shanghai 200444, People's Republic of China*

Inelastic neutron experiments were conducted on the cold-neutron triple-axis spectrometer Sika to study the spin-wave and crystal-field (CF) excitations from $ErFeO_3$. With the *ac* plane as the scattering plane, an excitation peak was observed at the low energy range below the ordering temperature $T_{N(\mathrm{Er})}$ of $Er^{3+}$ spins. Following this discovery, a series of energy scans in the range from 1.5 K to 10 K were performed at a off-centre position Q = (1 0 0.5). The results are plotted in Fig. S1. It is clearly seen that a low-energy excitation at ~ 0.35 meV at 10 K gradually shifted to a higher energy position ~ 0.7 meV upon cooling to ~1.5 K. A further experiment with higher energy resolution was carried out on the time-of-flight spectrometer Pelican, showing almost the same trend of the energy shift for this excitation. The higher resolution of Pelican allowed to resolve this excitation peak much better at all the temperatures (see Fig. 1 in the main article).

Further measurements with wide energy transfers were conducted on Sika to study the magnetic excitations in $ErFeO_3$ at low temperatures. The measurements mainly focused on the Q range in the vicinity of the antiferromagnetic (AFM) zone centre Q (1 0 1) of $ErFeO_3$ in the energy regime below 10 meV. Fig. S2 (a), (b) and (c) present the low-energy spectrum maps from the $ErFeO_3$ single crystal in this Q region at 1.5 K, 10 K and 150 K, respectively. There are two main discernible excitations in these figures. The first excitation is the spin-wave excitation at the zone centre, which is vertically distributed in a narrow Q space at low energy and slightly broadens at higher energy, indicating a sharp spin-wave dispersion, consistent with the previous reports in many orthoferrites. An energy gap is observed around 3 meV at both 1.5 K and 10 K for the spin-wave excitation. The spin-wave excitations at these two temperatures are very similar while the intensity was significantly enhanced when heating to 150K and the spin gap was suppressed, indicating the reduced magnetic single-ion anisotropy at high temperatures.

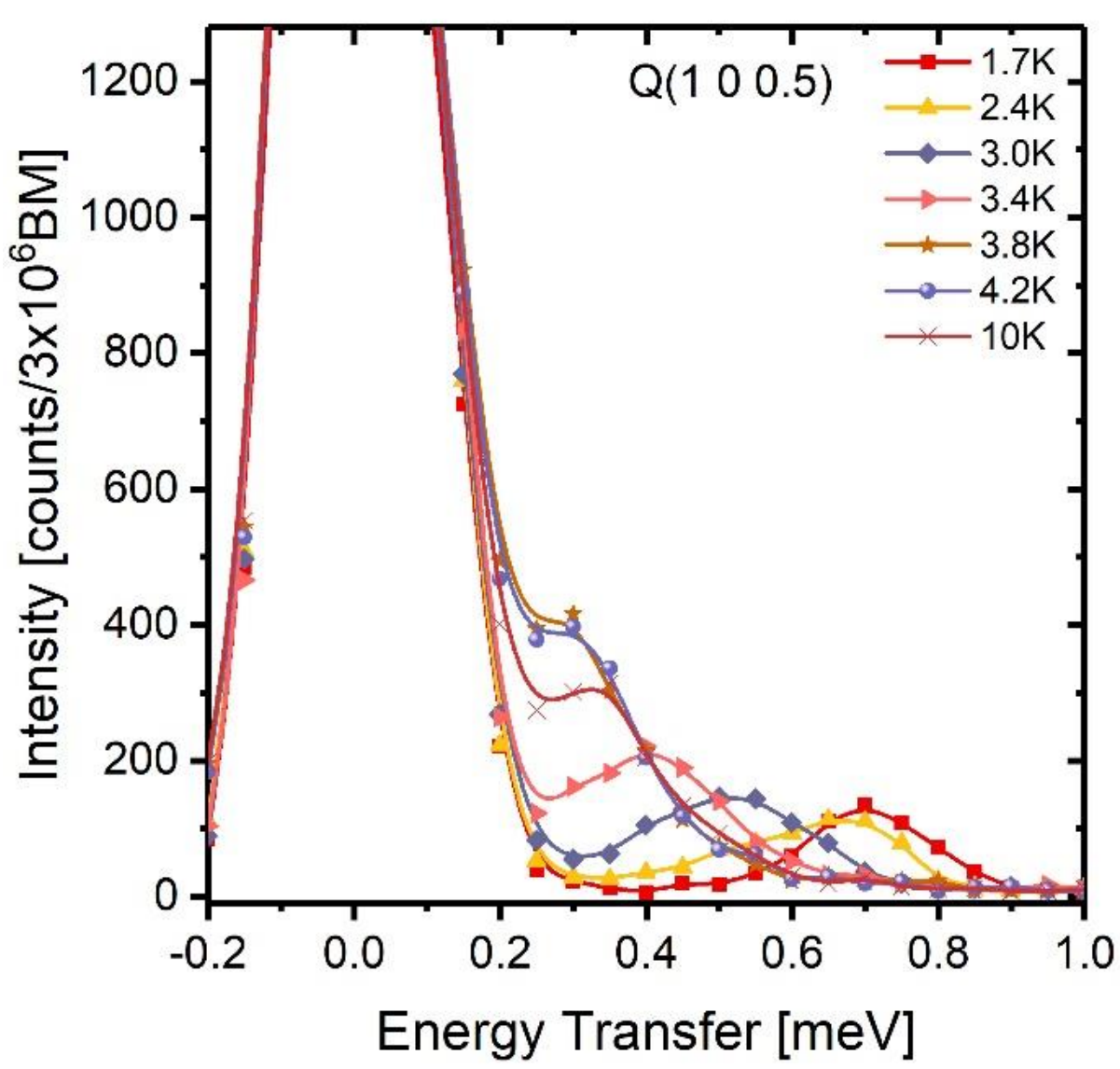

FIG. S1. Low-energy scans at Q (1 0 0.5) in $ErFeO_3$ at different temperatures on Sika. The symbol curves are the experimental data collected on Sika. The solid-line curves are the fitting to the experimental data by convoluting with the instrument resolution of Sika using the configuration of the experiment.

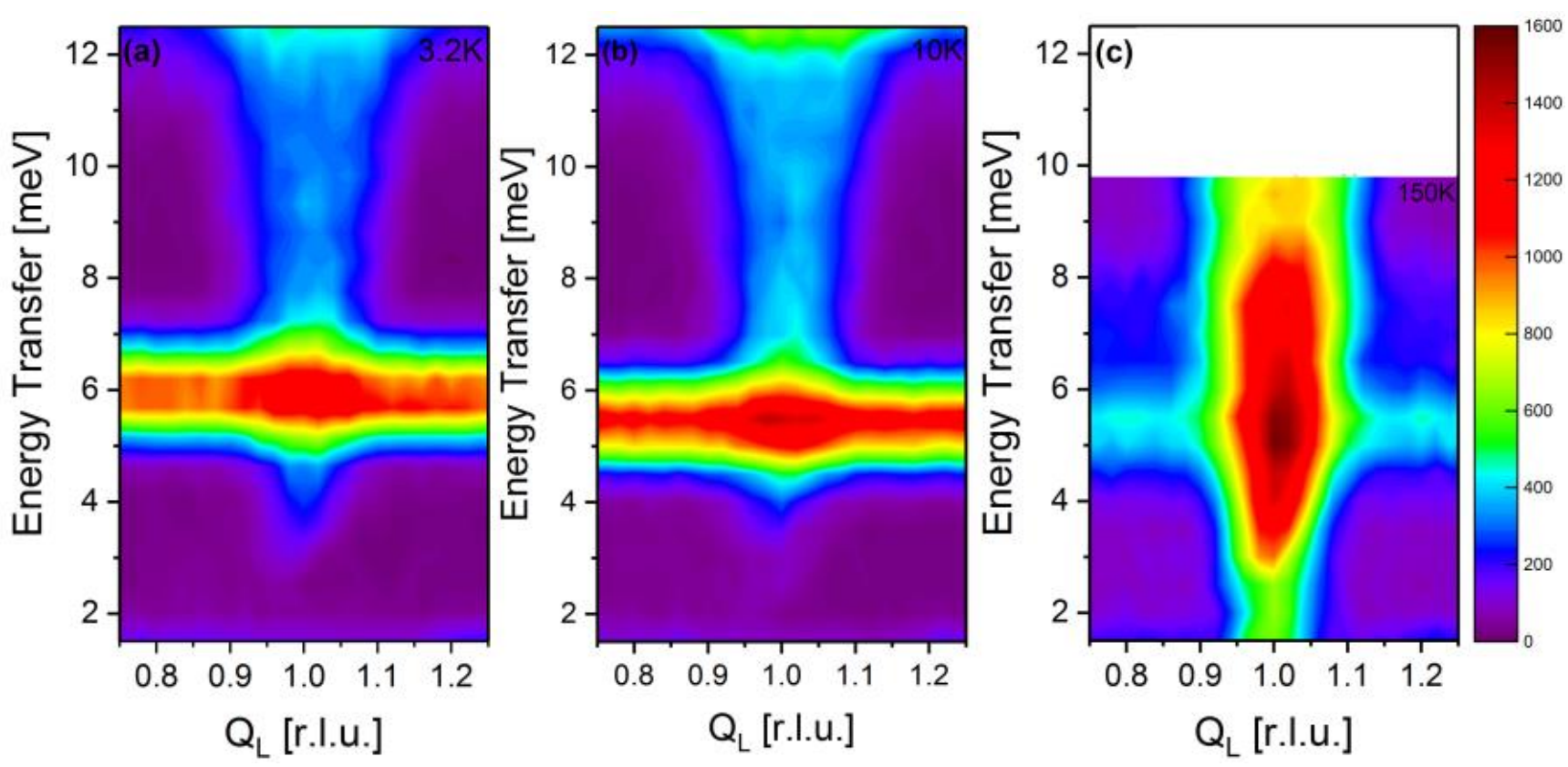

FIG. S2. False-colour contour maps of the spin gap and CF excitations of $ErFeO_3$ near the antiferromagnetic zone centre Q = (1 0 1) measured at (a) 1.5 K, (b) 10 K and (c) 150 K.

The second inelastic feature in Fig. S2 is the strong excitation at ~ 5.75 meV, which shows no dispersion along Q. It is safe for us to attribute such a non-dispersive peak to an CF excitation. Fig. S3 (a), (b) and (c) shows the two selected zone-centre and off-centre energy scans at 1.5 K, 90 K and 150 K, respectively. The peak intensity and shape evolutions of these two excitations can be clearly seen. It is found that the CF peak slightly shifted in energy upon heating. The excitation energy decreases from ~5.75 meV at 1.5 K to ~ 5.4 meV at 90 K, indicating the strong effect from the long-range ordering of $Er^{3+}$ magnetic moments. At the same time, the CF peak intensity drops significantly. A shoulder appeared on the high-energy side (~ 7.5 meV) of this CF peak. Since this peak was not observed at 1.5 K, we attributed it to the excitation from one excited CF level to another excited CF level at higher energy. See the main article for the discussion in detail.

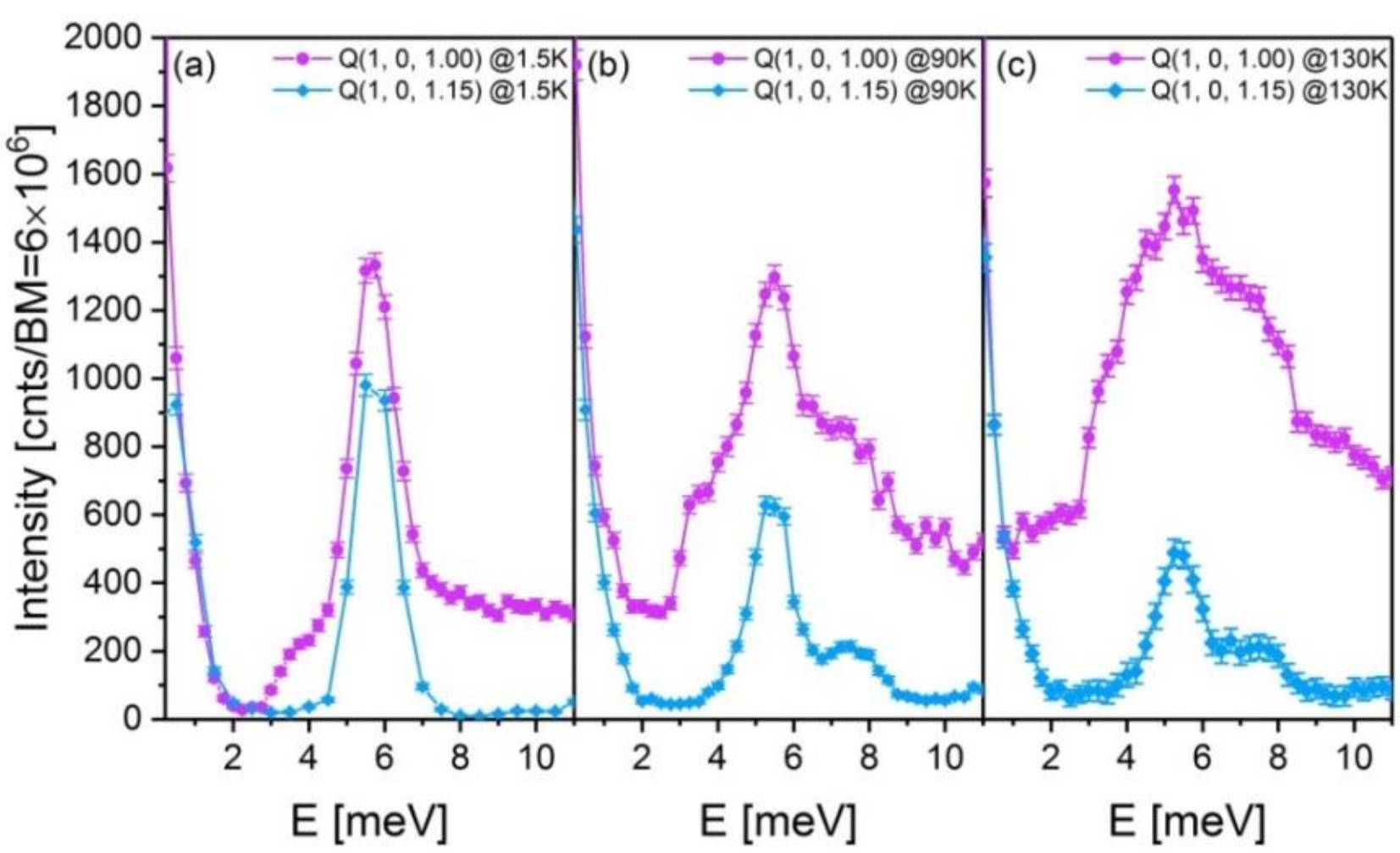

FIG. S3. Energy scans at the zone centre Q (1 0 1) and Q (1 0 1.15) at (a) 1.5 K, (b) 90 K and (c) 150 K. The purple curves correspond to the zone centre scans while the blue curves are the off-centre scans.

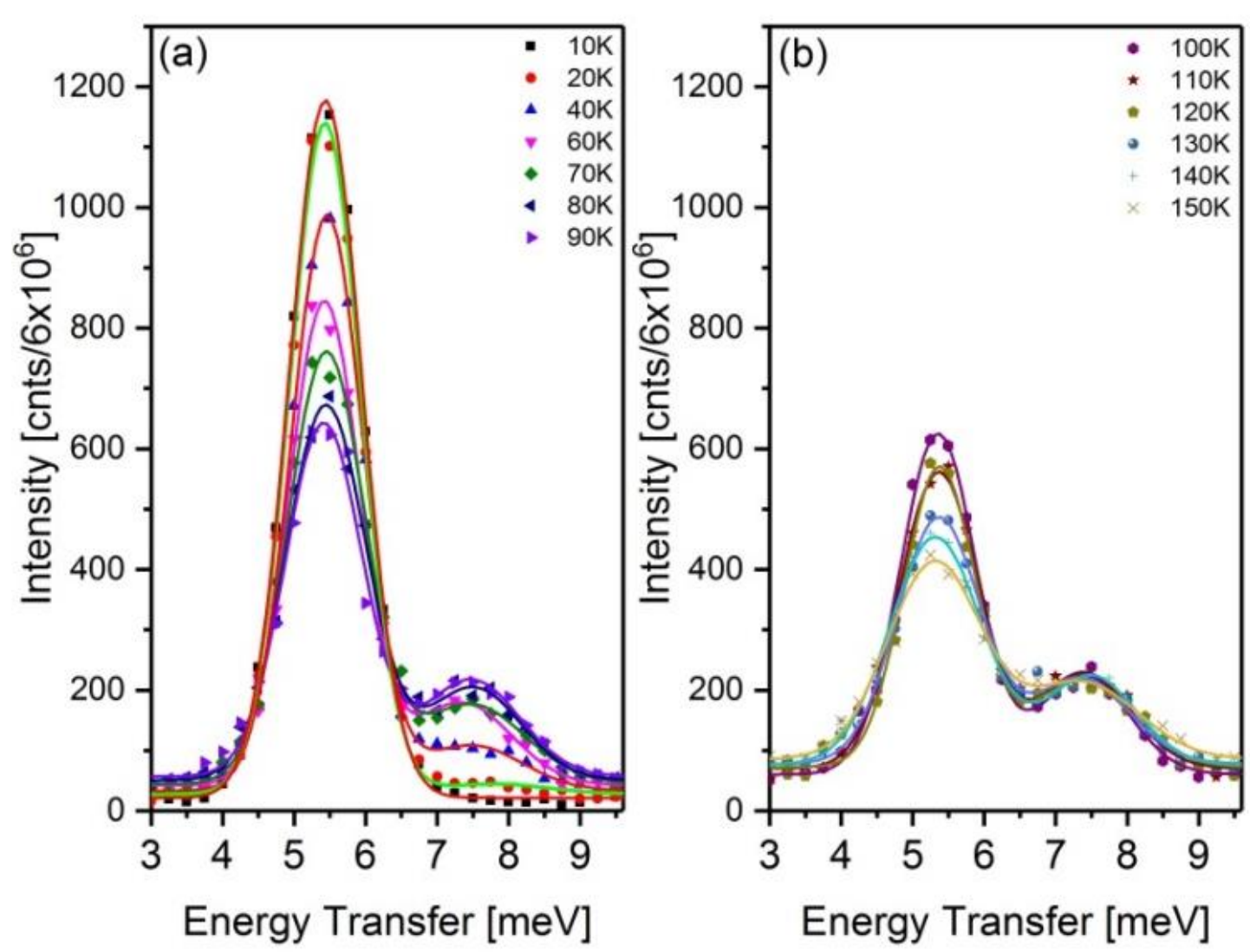

FIG. S4. Energy scans at the off-centre Q (1 0 1.15) position at different temperatures. The peaks in the data are fitted by convoluting with the instrument resolution of Sika. The main CF excitation peak at ~ 5.75 meV gradually decreases with the increase of temperature while the shoulder peak at ~ 7.5 meV shows up and gradually increases upon heating.

Fig. S4 (a) and (b) plots the off-centre energy scans at different temperatures with a fine temperature step. The decrease of the main peak at ~ 5.75 meV and the increase of the secondary peak at ~ 7.5 meV were clearly displayed. Fitting the data with two Lorentzians by convoluting with the instrument resolution, we obtained the fitted peak positions and intensities (see Fig. S5 (a) and (b)). Below $T_{N(Er)}$, the main excitation energy is at ~ 5.75 meV, much higher than those values observed at higher temperatures. This peak shows three steps in the full temperature range: ~ 5.75 meV below $T_{N(Er)}$, ~ 5.5 meV from $T_{N(Er)}$ to the low spin-reorientation temperature $T_l$, and ~ 5.35 meV above the upper spin-reorientation temperature $T_u$. The secondary peak at ~ 7.5 meV (purple curve) shows a smooth decreasing tendency upon heating from base to 150K. Its intensity gradually increases with the temperature rising from 1.5 K to 90 K, and almost saturates above 90 K. Such a behaviour indicates it as an excitation from one excited state to another higher-energy excited state. The sum of the two peak intensities (black curve) still follows a gradual decrease upon heating.

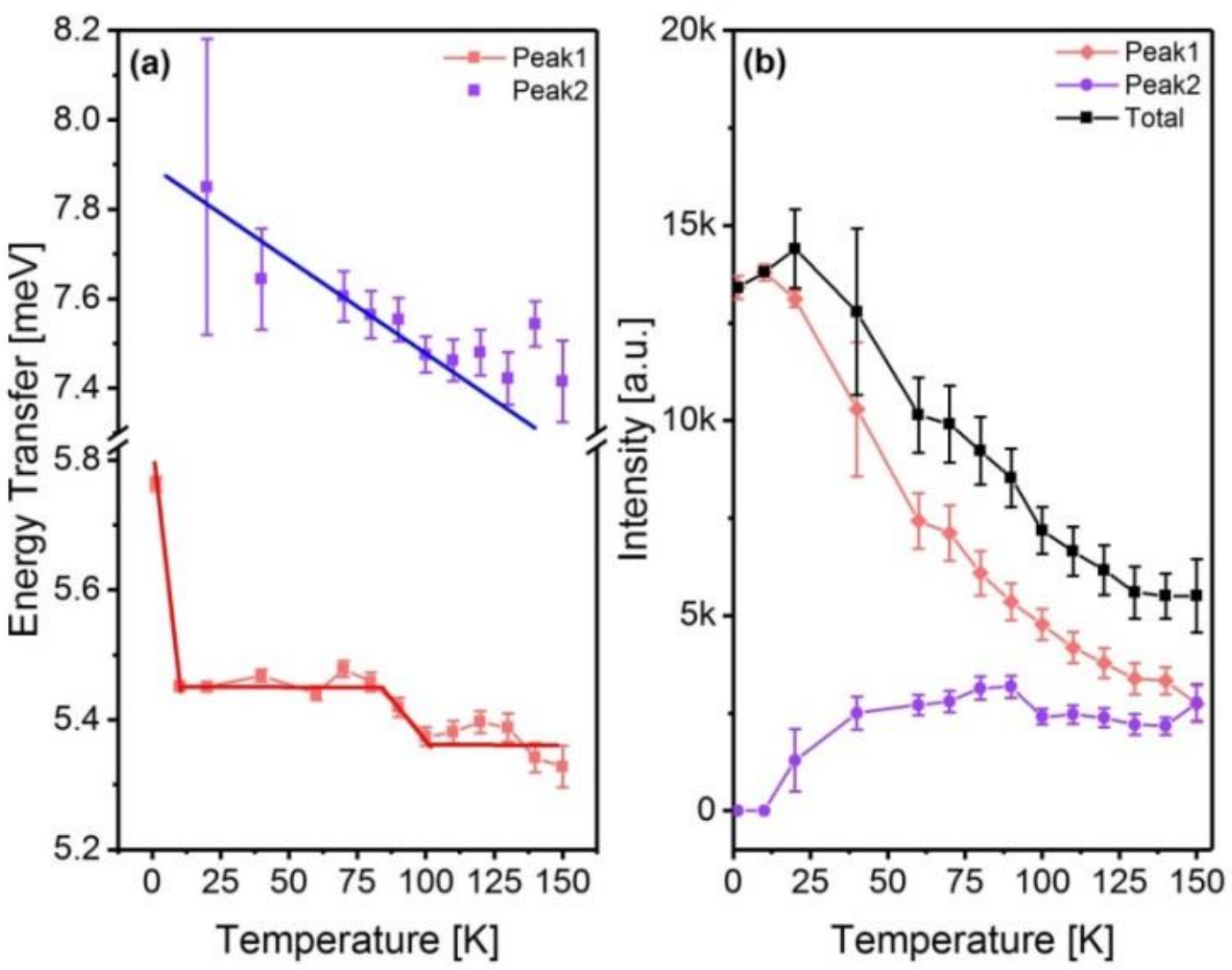

FIG. S5. The temperature dependency of the fitted peak positions (a) and intensities (b) from the energy scans in Fig. S4. The solid lines in (a) are guides to the eyes. The solid lines in (b) simply connect the data points.

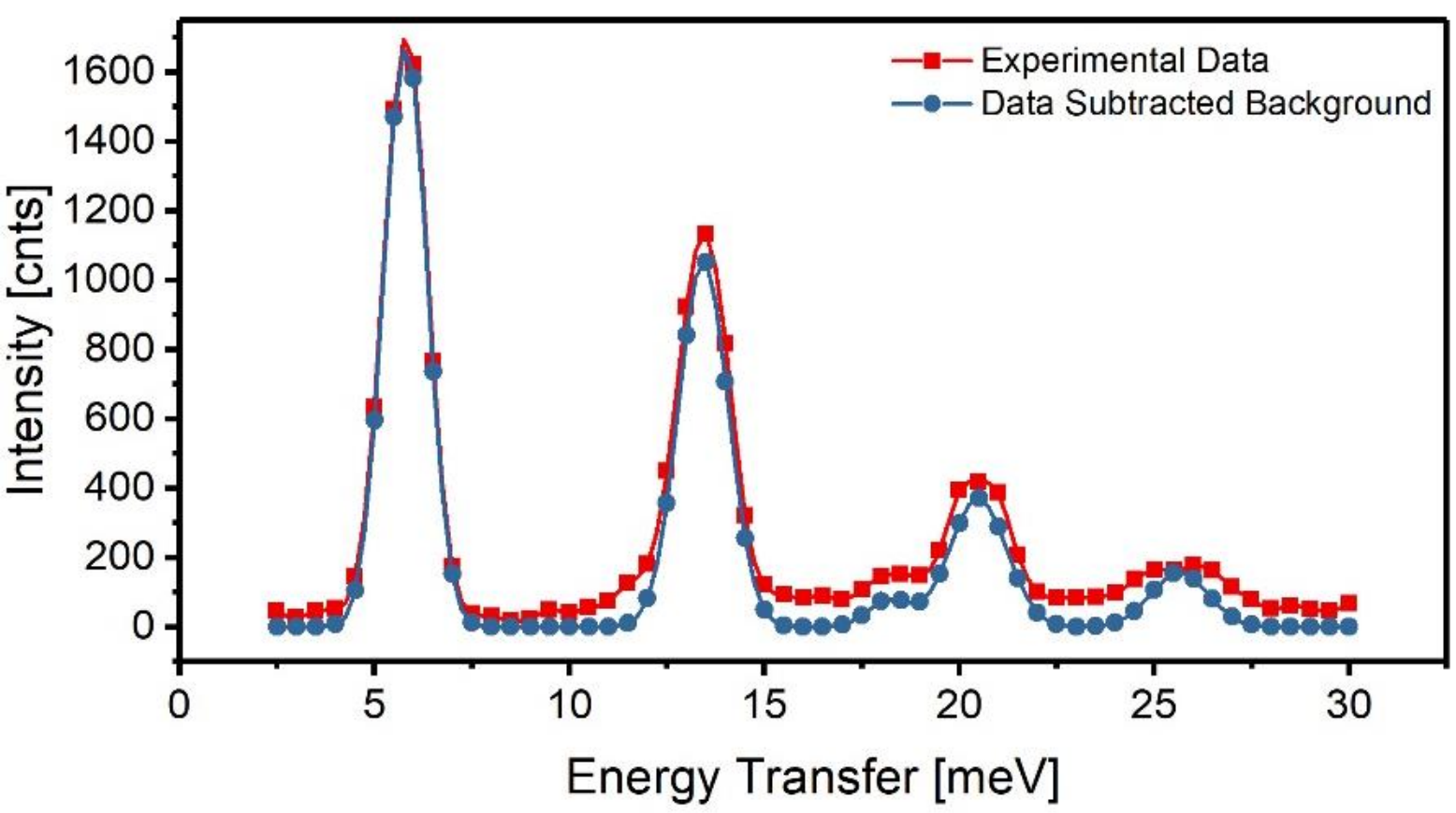

FIG. S6. The wide energy scan (red) from $ErFeO_3$ single crystal at 1.5K on Taipan. The peaks in the red curve are fitted to Voigt profile functions. The difference between the fitted peaks and the experimental data are subtracted from the original data as the background. The dark blue curve is the data with the subtracted background.

Fig. S6 shows the energy scan used for the CF model fitting described in the main article. Since the $Fe^{3+}$ magnetic ordered phase has a sharp spin-wave spectrum with the band top up to 60meV, it gets slightly broader at the high energy range. Thus, it generates a background to the off-centre scans which gradually increases in the high energy range in Fig. S6. This background was removed by the method described in the legend of the figure for the convenience to fit the data to the CF model.

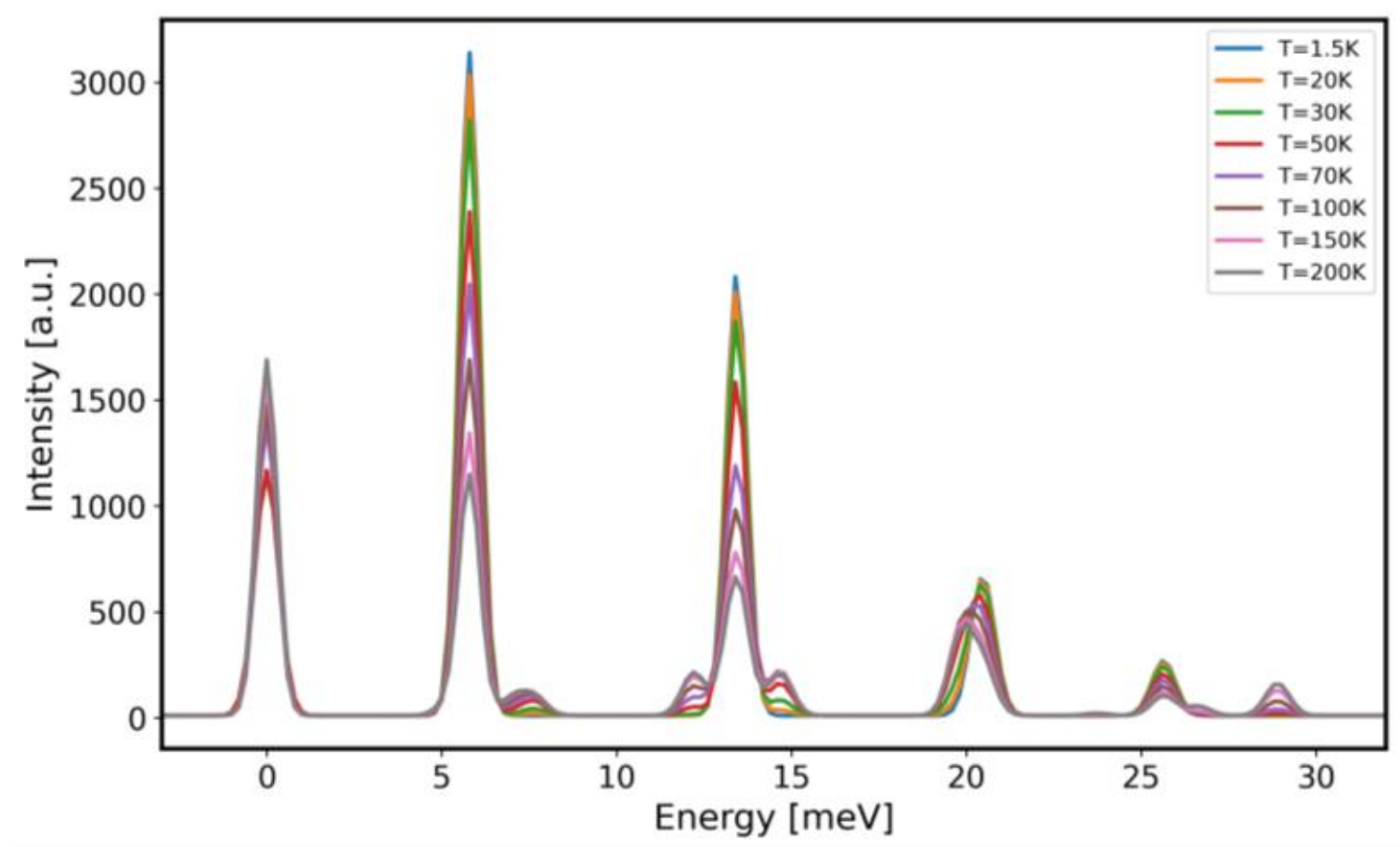

FIG. S7. The calculated CF excitation spectra at different temperatures. An additional peak appears above the peak at ~ 5.75 meV, consistent with the experimental data.

Fig. S7 display the simulated $Er^{3+}$ CF levels at different temperatures. One additional peak clearly shows up at ~ 7.5 meV when the sample is heated up from the base temperature. This new peak is consistent with the observation, which could be attributed to the excitation from the first excited level (~ 5.75 meV) to the second excited level (~ 13.5 meV). Two extra peaks were shown up on both sides of the peak at ~ 13.5 meV. They correspond to the excitation from the CF level at ~ 13.5 meV to the level at ~ 25 meV and the excitation from the level at ~ 5.75 meV to the third level at ~ 20.5 meV, respectively.